\documentstyle[psfig]{basi}
\begin{document}

\title[GMRT HI imaging of ERIDANUS]{GMRT HI imaging survey of ERIDANUS group of galaxies}
\author[A. Omar et al.]%
       {Amitesh Omar\thanks{e-mail:aomar@rri.res.in},  K.S. Dwarakanath, and K.R. Anantharamaiah\thanks{Deceased} \\
        Raman Research Institute, Bangalore 560 080}

\maketitle
\label{firstpage}

\begin{abstract} This project aims to image a subset of galaxies in the nearby group Eridanus in HI emission and in
radio continuum using the GMRT. The typical resolution of HI images will be $\sim$1~kpc with a 3$\sigma$ sensitivity to
HI mass of a few times 10$^{7}$ M$_\odot$.  The rotation curves obtained using the HI velocity field together with the
2MASS magnitudes in the near infrared will be used to construct the Tully-Fisher relation for this group. High
resolution (sub-kpc) multi-frequency radio continuum images will be used to study star-formation rates, radio
morphologies and spectral indices of Eridanus galaxies. About half of the galaxies have already been observed and
further observations on some late type galaxies are proposed with the GMRT. We present here the HI image and the
rotation curve for a member of this group viz. NGC~1385.  
\end{abstract}

\begin{keywords}
Galaxy -- HI: Galaxy -- ISM: group -- galaxy 
\end{keywords}
\section{Introduction}
Eridanus is a nearby group of galaxies at a distance of $\sim$20 Mpc. The group members are arranged in a filament like
structure joining Fornax and Dorado clusters of galaxies. Total number of optically identified galaxies in this group
is $\sim100$. This group has almost equal number of early type and late type systems. The redshift is known for all
galaxies through optical spectroscopy. The velocity dispersion of this group is $\sim350$ km~s$^{-1}$. The dynamical
study of this group shows that the group is a bound system comprising of several sub-groups \cite{wil89}. These
sub-groups are merging together and will eventually evolve to a massive cluster.

\section{Motivations for the present study}
The GMRT data of HI emission and radio continuum from the Eridanus galaxies, combined with the existing optical and 
infrared data will be used to study --

\noindent{1. Rotation curves of the  disk galaxies}\\
{2. Tully-Fisher relation using rotation curve and 2MASS magnitudes}\\
{3. M/L ratios and dark matter haloes of the disk galaxies}\\
{4. Star formation rates in the disk galaxies}\\
{5. Radio continuum morphology and spectral indices of galaxies}

\section{Results}
We have observed 26 pointings, each with a field of view of $\sim$24', using the GMRT and most of the data have been
analysed.  Further observations are proposed with the GMRT. The HI image and the rotation curve of NGC~1385, a member
of this group, are shown in figure~1.

\begin{figure}[h]
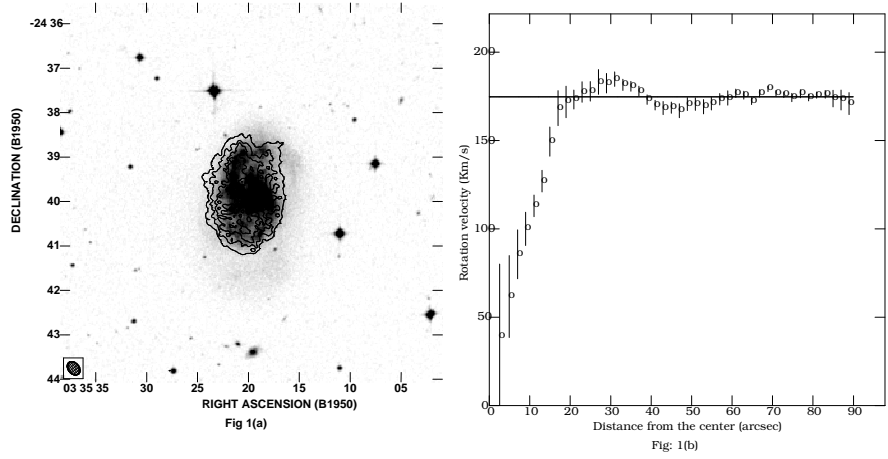

\begin{centering}
{\mbox {\psfig{file=aomar.fig1.ps,width=2.3truein,angle=0}}}
{\mbox {\psfig{file=aomar.fig2.ps,width=2.3truein,angle=0}}}
\caption{1(a). The HI column density contours of NGC 1385 are shown overlaid upon the DSS image. The contour levels
are 1, 3, 5, 7, 9 in units of $4.0\times10^{20}$ cm$^{-2}$. The HI mass is $2\times10^{9}$ M$_\odot$. The synthesised
beam ($20"\times14"$) shown in the lower left corner corresponds to a resolution of $\sim$1 kpc. 1(b). Rotation curve
of NGC 1385 constructed using the HI velocity field. The velocity resolution is $\sim13$ km s$^{-1}$. 1' corresponds 
to a linear distance of $\sim$5 kpc.} 
\end{centering}
\end{figure}

\label{lastpage}

\end{document}